\documentclass[12pt]{iopart}
\usepackage{iopams}
\usepackage{epsfig}

\begin{document}

\title{Enhancement of the upper critical field in the cubic Laves-phase superconductor HfV$_{2}$ by Nb doping}

\author{Jifeng Wu$^{1,2}$, Bin Liu$^{1,2}$, Yanwei Cui$^{4,2}$, Qinqing Zhu$^{1,2}$, Zhicheng Wang$^{4}$, Zhengwei Zhu$^{3}$, Hangdong Wang$^{5}$, Jian Li$^{2}$, Guanghan Cao$^{4}$, and Zhi Ren$^{2}$\footnote[1]{Electronic address: zhi.ren@wias.org.cn}}

\address{$^{1}$Department of Physics, Fudan University, Shanghai 200433, P. R. China}
\address{$^{2}$School of Science, Westlake Institute for Advanced Study, Westlake University, 18 Shilongshan Road, Hangzhou 310064, P. R. China}
\address{$^{3}$Wuhan National High Magnetic Field Center, School of Physics, Huazhong University of Science and Technology, Wuhan 430074, P. R. China}
\address{$^{4}$Department of Physics, Zhejiang University, Hangzhou 310027, P. R. China}
\address{$^{5}$Department of Physics, Hangzhou Normal University, Hangzhou 310036, P. R. China}

\date{\today}

\begin{abstract}
We report the effect of Nb doping on the upper critical field of the cubic Laves-phase superconductor HfV$_{2}$ studied in a series of HfV$_{2-x}$Nb$_{x}$ samples with 0 $\leq$ $x$ $\leq$ 0.3 under pulsed fields up to 30 T. The undoped HfV$_{2}$ undergoes a martensitic structural transition around 110 K, and becomes superconducting below $T_{\rm c}$ = 9.4 K. Upon Nb doping, while the structural transition is suppressed for $x$ $\geq$ 0.1, a maximum in $T_{\rm c}$ of 10.1 K and zero-temperature upper critical field $B_{\rm c2}$(0) of 22.4 T is found at $x$ = 0.2, which is ascribed to an increase of the density of states at the Fermi level. For all samples, the temperature dependence of $B_{\rm c2}$ can be well described by the Werthamer-Helfand-Hohenberg (WHH) theory that takes into account both the spin paramagnetic effect and spin orbit scattering. In addition, a comparison is made between the $B_{\rm c2}$ behavior of HfV$_{2-x}$Nb$_{x}$ and those of Nb-Ti and Nb$_{3}$Sn.
\end{abstract}
\maketitle

\section{\label{sec:level1}Introduction}
HfV$_{2}$ based superconductors with the C15 cubic Laves-phase structure have received considerable attention because of their potential as high-field superconducting magnets \cite{HfV2HF1,HfV2HF2}.
Similar to A15-type Nb$_{3}$Sn, HfV$_{2}$ undergoes a martensitic transition involving two successive structural modifications from cubic to tetragonal then to orthorhombic in the temperature range between 100 to 115 K \cite{HfV2structure}, followed by a superconducting transition below $T_{\rm c}$ $\sim$ 9 K \cite{SCinHfV2}.
Despite its relatively low $T_{\rm c}$, the upper critical field $B_{\rm c2}$ of HfV$_{2}$ has been found to be $\sim$20 T at 4.2 K, and can be improved to $\sim$23 T by alloying with ZrV$_{2}$ \cite{Hf1-xZrxV2}, which is the highest for superconductors composed only of transition metal elements.
Furthermore, compared with Nb$_{3}$Sn, the HfV$_{2}$-based C15 compounds are more resistant to neutron radiation \cite{HfV2neutron} and less brittle \cite{HfV2plastic},
which make them promising for application in nuclear fusion reactor.

For superconductors with spin-singlet Cooper pairs, superconductivity(SC) can be destroyed by the application of magnetic field via two effects.
The first one is the orbit effect, which is a manifestation of the Lorentz force. The second one is the spin paramagnetic effect, which tends to align the spin of Cooper pairs and can be partially cancelled out by introducing the spin-orbit scattering. The investigation of the $B_{\rm c2}$ has been shown to provide valuable information on the contributions from these effects \cite{WHHtheory1,Makiparameter,WHHtheory2}.
In the case of Hf$_{1-x}$Zr$_{x}$V$_{2}$, the $B_{\rm c2}$ value was measured across 0 $\leq$ $x$ $\leq$ 1, by only at 4.2 K \cite{Hf1-xZrxV2}.
In addition to Zr, various other dopants, such as Ta \cite{HfV2Tadoping}, Nb \cite{HfV2Nbdoping}, Fe \cite{HfV2FeCoNidoping}, Co \cite{HfV2FeCoNidoping}, Ni \cite{HfV2FeCoNidoping}, have been introduced in the HfV$_{2}$ system to investigate the impurity effect on superconductivity and structural transition.
For example, dilute Nb doping ($\leq$ 5\%) is found to suppress the structural transition and increase $T_{\rm c}$ \cite{HfV2Nbdoping}.
However, no systematic study on the temperature and doping dependencies of $B_{\rm c2}$ has been reported for any of these dopants.

In this paper, we present a study on the superconducting properties of HfV$_{2-x}$Nb$_{x}$ for $x$ in the range of 0 to 0.3, focusing on the $B_{\rm c2}$($T$) behavior. Bulk SC with indication of strong coupling nature is observed in this $x$ range, and a maximum $T_{\rm c}$ of 10.1 K is found at $x$ = 0.2.  The $B_{\rm c2}$($T$) data for all $x$ values are well fitted by the WHH model including the effects of spin paramagnetism and spin orbit scattering. The extrapolated $B_{\rm c2}$(0) exhibits a maximum of 22.4 T also at $x$ = 0.2, which is 4.1 T higher than that of undoped HfV$_{2}$.
The enhancement of both $T_{\rm c}$ and $B_{\rm c2}$(0) is attributed to an enhanced density of states of the Fermi level [$N(E_{\rm F})$], as revealed by the specific heat results.
We also compare the $B_{\rm c2}$($T$) curve at the optimal doping with those of Nb$_{0.44}$Ti$_{0.56}$ and Nb$_{3}$Sn, and discuss its implication on the $B_{\rm c2}$ improvement.

\section{\label{sec:level1}Experimental details}
Polycrystalline HfV$_{2-x}$Nb$_{x}$ samples with $x$ = 0, 0.1, 0.2, 0.3 were prepared by the arc-melting method. High purity Hf slugs ($>$99.9\%), V ($>$99.99\%, 200 mesh) and Nb ($>$99.99\%, 200 mesh) powders were weighed according to the stoichiometric ratio and melted in an arc furnace under high-purity argon atmosphere. The melts were turned over and remelted by the same procedure several times to ensure homogeneity, following by rapid cooling on a water-chilled copper plate.
The samples were used as cast without further annealing treatment, which was reported to result in $T_{\rm c}$ degradation \cite{HfV2Tcdegradation}. The phase purity of the samples was checked by powder x-ray diffraction (XRD) using a Bruker D8 Advance x-ray diffractometer with Cu K$\alpha$ radiation at room temperature.
The chemical composition of theses samples was examined with an energy-dispersive x-ray (EDX) spectrometer (Model Octane Plus) affiliated to a Zeiss Supratm 55 schottky field emission scanning electron microscope (SEM). The spectra were collected on at least 3 different locations of each sample for averaging. The electrical resistivity was measured using a standard four-probe method. Resistivity and specific heat measurements down to 1.8 K and up to 9 T were carried out on regular-shaped samples in a Quantum Design PPMS-9 Dynacool.
The magnetoresistance under pulsed fields up to 30 T was measured at the National High Magnetic Field Center in Wuhan. The dc magnetization measurements down to 1.8 K were performed using a commercial SQUID magnetometer (Quantum Desgin MPMS3).

\section{\label{sec:level1}Results and Discussion}
Figure 1(a) shows the room-temperature XRD patterns for the HfV$_{2-x}$Nb$_{x}$ samples.
All the major peaks can be indexed on the cubic C15 structure with the space group Fd3m \cite{HfV2structure}, and a schematic structure is shown in Fig. 1(b).
The lattice parameter, determined by a least-squares method, is plotted as a function of the nominal Nb content $x$ in Fig. 1(c). For undoped HfV$_{2}$ ($x$ = 0), the refined $a$ = 7.382 {\AA} is in good agreement with the previous reports \cite{HfV2structure,SCinHfV2}. In this compound, the V atoms form a network of corner sharing tetrahedron, and the nearest V-V distance $d_{\rm V-V}$ = 2.616 {\AA}  is even smaller than that (2.620 {\AA}) in body-center-cubic (BCC) V.
With increasing $x$, the $a$-axis expands monotonically to 7.399 {\AA}. This is in line with the expectation that Nb substitutes V rather than Hf since $r$(V) $<$ $r$(Nb)$<$ $r$(Hf), where $r$ is the atomic radius.
In addition to the main C15 phase, there exist small extra diffraction peaks, which is ascribed to the Hf-based solid solution impurity. Base on the peak intensity at $\sim$36.7$^{\circ}$ (marked by the asterisks),
the impurity fraction is estimated to vary from $\sim$8\% to $\sim$12\%, depending on the Nb content.
On the other hand, the chemical compositions of these samples measured by EDX are HfV$_{2.04(5)}$, HfV$_{1.88(3)}$Nb$_{0.12(2)}$, HfV$_{1.7(1)}$Nb$_{0.20(4)}$, and HfV$_{1.67(2)}$Nb$_{0.35(3)}$, which agree with the nominal ones considering both the measurement error and impurity level.

Figure 2(a) shows the low temperature resistivity for the HfV$_{2-x}$Nb$_{x}$ samples. A resistive superconducting transition is observed for all the samples.
For HfV$_{2}$ ($x$ = 0), the transition starts at $\sim$9.6 K and the transition width is about 0.4 K.
In addition, as shown in the inset of Fig. 2(a), a resistivity bump due to the structural phase transition is clearly visible at $\sim$110 K.
Upon Nb doping, the resistive transition sharpens considerably.
The $T_{\rm c}$ values, determined as midpoint temperatures of the resistive transitions, are 9.4 K, 9.88 K, 10.1 K, and 9.84 K for $x$ = 0, 0.1, 0.2, and 0.3, respectively.
The occurrence of SC in these HfV$_{2-x}$Nb$_{x}$ samples is corroborated by the magnetic susceptibility results shown in Fig. 2(b).
For each $x$, a strong diamagnetic signal is detected, and its onset temperature corresponds well to the midpoint of the resistive transition except for $x$ = 0, where the onset temperature coincides with the completion of the resistive transition. When cooling below 9 K, all the zero-field cooling susceptibility curves become flat, and the shielding fractions are calculated to vary between 110\% and 150\% without correction for the demagnetization effect.

The main panel of Fig. 3(a) shows the temperature dependence of the specific heat $C_{\rm p}$/$T$ for the HfV$_{2-x}$Nb$_{x}$ samples.
A $C_{\rm p}$ anomaly can be seen at $\sim$110 K for HfV$_{2}$, in accordance with the martensitic structural transformation.
However, the small magnitude and relatively broad width of this anomaly signify that the structural transition is of second order rather than first order \cite{HfV2Nbdoping}.
This is probably because that our sample is slightly off stoichiometric, and, indeed, its $T_{\rm M}$ $\approx$ 110 K is slightly lower than that the usual value of 116 K \cite{HfV2Nbdoping}.
For $x$ $\geq$ 0.1, no such anomaly is discernible. This indicates that the structural transformation is suppressed completely by this Nb doping level, in line with the previous study \cite{HfV2Nbdoping}.
On the other hand, a sharp $C_{\rm p}$ jump around $T_{\rm c}$ is found for all samples, which can be seen more clearly from the $C_{\rm p}$/$T$ versus $T^{2}$ plot shown in the inset of Fig. 3(a).
This confirms the bulk nature of SC in these HfV$_{2-x}$Nb$_{x}$ samples.
Following Ref. \cite{HfV2Nbdoping}, the normal-state data are analyzed by the Debye model $C_{\rm p}$/$T$ = $\gamma_{\rm e}$ + $\beta_{3}$$T^{2}$ + $\beta_{5}$$T^{4}$ + $\beta_{7}$$T^{6}$, where $\gamma_{\rm e}$ and $\beta_{i}$ ($i$ = 3, 5, 7) are the electronic and phonon specific heat coefficients, respectively. Here the high order terms are included since there is a downward curvature of the data.
The best fits yield $\gamma_{\rm e}$ = 50 mJmol$^{-1}$K$^{-2}$, 65 mJmol$^{-1}$K$^{-2}$, 68 mJmol$^{-1}$K$^{-2}$, and 62 mJmol$^{-1}$K$^{-2}$ for $x$ = 0, 0.1, 0.2, and 0.3, respectively,
which are comparable to the previous study \cite{HfV2Nbdoping}.
Since $\gamma_{\rm e}$ is proportional to $N(E_{\rm F})$, these results not only indicate a large $N(E_{\rm F})$ in HfV$_{2-x}$Nb$_{x}$\cite{HfV2largeNEF}, but also point to an increase of $N(E_{\rm F})$ by Nb doping.

Figure 3(b) shows the temperature dependence of the normalized electronic specific heat $C_{\rm el}$/$\gamma_{\rm e}$$T$ after subtraction of the phonon contribution. The normalized specific heat jumps $\Delta$$C_{\rm el}$/$\gamma_{\rm e}$$T_{\rm c}$ for these samples are about 1.7, which are significantly larger than the value 1.43 of the Bardeen-Cooper-Schrieffer (BCS) theory \cite{BCStheory} and point to a strong coupling superconducting state. In such case, the $C_{\rm el}$/$\gamma_{\rm e}$$T$ jump can be analyzed by a modified BCS model, or the so called "$\alpha$ model",
where $\alpha$ = $\Delta_{0}$/$T_{c}$ and $\Delta_{0}$ is the size of fully isotropic gap at 0 K \cite{alphamodel}.
It turn out that the data of $x$ = 0 can be well reproduced with $\Delta_{0}$/$T_{c}$ = 2, which is very close to that reported previously \cite{HfV2specificheat} and indeed larger than the BCS value of 1.764. For Nb-doped samples, the data almost overlap with each other and can be better fitted with $\Delta_{0}$/$T_{c}$ = 2.25 for $T$/$T_{\rm c}$ $\leq$ 0.6. Nevertheless, at higher $T$/$T_{\rm c}$, a discrepancy exists between the experimental and theoretical results, which is probably due to either the error in the estimation of phonon specific heat contribution or the existence of gap anisotropy.

To obtain $B_{\rm c2}$($T$) for the HfV$_{2-x}$Nb$_{x}$ samples, temperature dependent resistivity measurements at constant fields up 9 T (in PPMS) are combined with isothermal magnetoresistance measurements under pulsed fields up to 30 T.
Examples for $x$ = 0.2 are shown in Fig. 4(a) and (b), respectively, and the suppression of SC with increasing field is evident in both cases.
For consistency, the value of $T_{c}$ ($B_{\rm c2}$) in each curve is determined as the onset temperature (field) of the resistive transition, and
the resulting $B_{\rm c2}$$-$$T$ phase diagrams are displayed in Fig. 4(c).
The initial slopes of $B_{\rm c2}(T)$, ($d$$B_{\rm c2}/dT$)$_{T = T_{\rm c}}$, are found to be $-$5.1 T/K, $-$6.0 T/K, $-$6.0 T/K, and $-$5.4 T/K for $x$ = 0, 0.1, 0.2, and 0.3, respectively, which are typical for HfV$_{2}$ based superconductors \cite{HfV2specificheat}.

According to the WHH theory \cite{WHHtheory1}, the zero-temperature orbital upper critical field $B_{\rm c2}^{\rm orb}$(0) in the dirty limit is given by
\begin{equation}
B_{\rm c2}^{\rm orb}(0) = -0.693(dB_{\rm c2}/dT)_{T = T_{\rm c}}T_{\rm c}.
\end{equation}
Using this formula, the calculated $B_{\rm c2}^{\rm orb}$(0) for the HfV$_{2-x}$Nb$_{x}$ samples falls between 33.2 to 41.9 T, which is much higher than the estimated $B_{\rm c2}$(0) $\sim$ 20 T.
This indicates that the spin paramagnetic effect is at work, and hence the Cooper pairs are in the spin singlet state.
Note that the paramagnetically limiting field $B_{\rm c2}^{\rm P}$(0) can be expressed as
\begin{equation}
B_{\rm c2}^{\rm P}(0) = B_{\rm c2}^{\rm orb}(0)/\sqrt{1 + \alpha_{\rm M}^{2}},
\end{equation}
where $\alpha_{\rm M}$ = $\sqrt{2}$$B_{\rm c2}^{\rm orb}$(0)/$B_{\rm P}$(0) is the Maki parameter and $B_{\rm P}$(0) is the Pauli limiting field at 0 K \cite{Makiparameter}.
Since $B_{\rm P}$(0) = $\Delta_{0}$/$\sqrt{2}$$\mu_{\rm B}$, where $\mu_{\rm B}$ is the Bohr magneton, the $\alpha_{\rm M}$ values are found to be 2.38, 2.47, 2.47, and 2.31 for $x$ = 0, 0.1, 0.2 and 0.3, respectively.
Then $B_{\rm c2}^{\rm P}$(0) are estimated to be in the range from 12.9$-$15.7 T, significantly smaller than the experimental values.
Clearly, spin-orbit scattering that counteracts the spin paramagnetic effect needs also to be considered.

It is known that the $B_{\rm c2}$($T$) of a type-II superconductor in the dirty limit can be calculated from the the linearized Gor'kov equation, with spin paramagnetic effect and spin-orbit scattering parameterized as $\alpha_{\rm M}$ and $\lambda_{\rm so}$ \cite{WHHtheory2}. The equation is expressed by a sum of digamma functions as:
\begin{equation}
\fl \rm ln\frac{1}{t} = \left(\frac{1}{2}+\frac{i\lambda_{\rm so}}{4\gamma}\right)\psi\left(\frac{1}{2}+\frac{\bar{h}+\lambda_{\rm so}/2+i\gamma}{2t}\right) + \left(\frac{1}{2}-\frac{i\lambda_{\rm so}}{4\gamma}\right)\psi\left(\frac{1}{2}+\frac{\bar{h}+\lambda_{\rm so}/2-i\gamma}{2t}\right) - \psi\left(\frac{1}{2}\right),
\end{equation}
where $t$ = $T/T_{\rm c}$, $\gamma$ $\equiv$ [($\alpha_{\rm M}$$\bar{h}$)$^{2}$-($\lambda_{\rm so}$/2)$^{2}$]$^{1/2}$, and
\begin{equation}
\bar{h} = \frac{4B_{\rm c2}}{\pi^{2}(-dB_{\rm c2}/dt)_{t=1}}
\end{equation}
The normalized upper critical field $h^{\ast}$ = ($\pi^{2}$/4)$\bar{h}$ for the HfV$_{2-x}$Nb$_{x}$ samples with $x$ $\leq$ 0.3 is plotted against $t$ in Fig. 5. As can be seen, all the data points are well fitted only when both $\alpha_{\rm M}$ and $\lambda_{\rm so}$ are nonzero, which are summarized in Table 1.
Actually, the theoretical curves concerning only the orbital effect ($\alpha_{\rm M}$ = 0, $\lambda_{\rm so}$ = 0) fit the data close to $T_{\rm c}$ well but strongly overestimate $B_{\rm c2}$($T$) at low temperature.
By contrast, the fitting curves with $\lambda_{\rm so}$ = 0 underestimate the $B_{\rm c2}$($T$) data for most of the $t$ range. These results not only demonstrate conclusively the presence of spin orbit scattering, but also underline the importance of low temperature data in the analysis of $B_{\rm c2}(T)$ behavior for HfV$_{2}$ based superconductors \cite{HfV2specificheat}.

To gain more insight, the values of $T_{\rm c}$, $B_{\rm c2}$(0), $\gamma_{\rm e}$, $\alpha_{\rm M}$ and $\lambda_{\rm so}$ from the above results are plotted as a function of the Nb content in Fig. 6.
It turns out that all of $T_{\rm c}$, $B_{\rm c2}$(0) and $\gamma_{\rm e}$ follow a nonmonotonic $x$ dependence with a maximum at $x$ = 0.2, hinting at a close relation between them, while no obvious feature is seen in the data of $\alpha_{\rm M}$ and $\lambda_{\rm so}$.
In particular, compared with undoped HfV$_{2}$, the $B_{\rm c2}$(0) value is 4.1 T ($\sim$22\%) larger at $x$ = 0.2, which is remarkable since $T_{\rm c}$ is increased only by 0.7 K ($\sim$7\%).
For strong coupling superconductors, their $T_{\rm c}$ is given by the MaMillan formula \cite{McMillan}
\begin{equation}
T_{\rm c} = \frac{\Theta_{\rm D}}{1.45}\rm exp\left[-\frac{1.04(1+\lambda)}{\lambda-\mu^{\ast}(1+0.62\lambda)]}\right],
\end{equation}
with
\begin{equation}
\lambda = \frac{N(E_{\rm F})\langle I^{2}\rangle}{M\langle \omega^{2}\rangle},
\end{equation}
where $\Theta_{\rm D}$ is the Debye temperature, $\mu^{\ast}$ is the Coulomb pseudopotential, $\langle I^{2}\rangle$ is the average square electron-ion matric element, $M$ is the atomic mass, and $\langle \omega^{2}\rangle$ is the average square phonon frequency.
Hence one could see that $T_{c}$ increases with increasing $N(E_{\rm F}$), which explains the enhancement of $T_{\rm c}$ by Nb doping.
On the other hand, it is noted from Fig.5 that the $h^{\ast}$ values at 0 K are very similar for the HfV$_{2-x}$Nb$_{x}$ samples.
It thus appears that $B_{\rm c2}$(0) is nearly proportional to the product of $T_{\rm c}$ and ($d$$B_{\rm c2}/dT$)$_{T = T_{\rm c}}$.
In the dirty limit, we have ($d$$B_{\rm c2}/dT$)$_{T = T_{\rm c}}$ $\propto$ $\gamma_{\rm e}$$\rho_{\rm c}$, where $\rho_{\rm c}$ is the resistivity value just above $T_{\rm c}$ \cite{dirtylimit}. Given that $\gamma_{\rm e}$ increases by more than 30\% with increasing $x$ from 0 to 0.2, it can be speculated that the concomitant enhancement of $B_{\rm c2}$(0) is also mainly due to an enhanced $N(E_{\rm F}$).

Finally, we present in Fig. 7 a comparison between the $B_{\rm c2}$-$T$ phase diagrams of HfV$_{1.8}$Nb$_{0.2}$, Nb$_{0.44}$Ti$_{0.56}$ \cite{Nb-Ti}, and Nb$_{3}$Sn \cite{Nb3Sn},
all of which adopt the cubic structure but with different space groups (Table 2).
It is noted that the $B_{\rm c2}$($T$) curve of HfV$_{1.8}$Nb$_{0.2}$ lies between those of Nb$_{0.44}$Ti$_{0.56}$ and Nb$_{3}$Sn.
In particular, the $B_{\rm c2}$(0) value of HfV$_{1.8}$Nb$_{0.2}$ is much higher than that of Nb$_{0.44}$Ti$_{0.56}$, despite their similar $T_{\rm c}$.
This substantiates that C15 HfV$_{2}$-based superconductors hold the record $B_{\rm c2}$(0) for compounds composed only of transition metal elements.
Furthermore, as can be seen from Table 2, $B_{\rm c2}$(0) already approaches its orbital-limited value for Nb$_{0.44}$Ti$_{0.56}$ and Nb$_{3}$Sn.
By contrast, given its relatively small $\lambda_{\rm so}$, there is still plenty room for the $B_{\rm c2}$(0) enhancement of HfV$_{1.8}$Nb$_{0.2}$.
Indeed, assuming the same $\lambda_{\rm so}$ = 4.5 as Nb$_{0.44}$Ti$_{0.56}$, $B_{\rm c2}$(0) of HfV$_{1.8}$Nb$_{0.2}$ would reach $\sim$33 T, which is even higher than that of Nb$_{3}$Sn.
In this respect, it is worthy noting the following two relations: $\lambda_{\rm so}$ = 2$\hbar$/3$\pi$$k_{\rm B}$$T_{\rm c}$$\tau_{\rm so}$ \cite{WHHtheory2} and $\tau_{0}$/$\tau_{\rm so}$ = ($Z$/137)$^{4}$ \cite{Abrisokov},
where $\tau_{\rm so}$ is the spin orbit scattering time, $\tau_{0}$ is the transport lifetime and $Z$ is the atomic number.
Hence, doping with high $Z$ elements, such as 4$d$ or 5$d$ elements, may increase $\lambda_{\rm so}$ and will be of interest for future studies.

\section{\label{sec:level1}Conclusion}
In summary, we have studied the superconducting properties, with focus on the upper critical field, of polycrytalline C15-type Laves phase compound HfV$_{2-x}$Nb$_{x}$ with $x$ up to 0.3.
While the martensitic structural transition is suppressed for $x$ $\geq$ 0.1, a maximum in $T_{\rm c}$ (= 10.1 K) as well as $B_{\rm c2}$(0) (= 23.1 T) is found at $x$ = 0.2.
Furthermore, the specific heat results indicate a strong coupling SC in the whole $x$ range studied, and an increase of $N(E_{\rm F}$) induced by Nb doping, which is suggested to be responsible for the enhancement of both $T_{\rm c}$ and $B_{\rm c2}$(0).
The $B_{\rm c2}$($T$) data of all samples are found to be well described by the WHH model considering the spin paramagnetic effect and spin orbit scattering.
In addition, a comparison between the $B_{\rm c2}$($T$) behavior of HfV$_{2-x}$Nb$_{x}$ and those of Nb-Ti and Nb$_{3}$Sn signifies the importance of spin orbit scattering in the $B_{\rm c2}$(0) improvement for the former case.
Our results call for further studies to increase the density of states at the Fermi level and spin orbit scattering in the HfV$_{2}$ based Laves-phase superconductors, which may facilitate the high-field application of these materials.

\section*{Acknowledgments}
We thank Dr. Yong Sun for his assistance in the data fitting. The work at Zhejiang University is supported by the National Key Research and Development Program of China (No.2017YFA0303002) and the Fundamental Research Funds for the Central Universities of China.

\pagebreak[4]

\begin{figure}
\centering
\includegraphics[width=12cm]{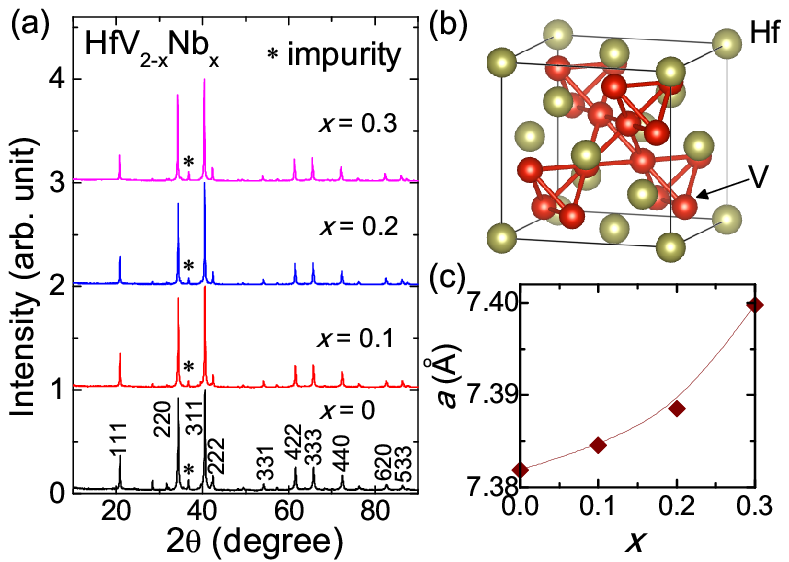}
\caption{\label{Fig.1}  (a) Powder x-ray diffraction patterns at room temperature for the HfV$_{2-x}$Nb$_{x}$ samples. The peaks for $x$ = 0 are indexed to the Fd3m space group.
The asterisks mark the impurity phase.
(b) Schematic structure of HfV$_{2}$.
(c) Cubic lattice parameter plotted as a function of the Nb content $x$.}
\end{figure}

\begin{figure}[h]
\centering
\includegraphics[width=12cm]{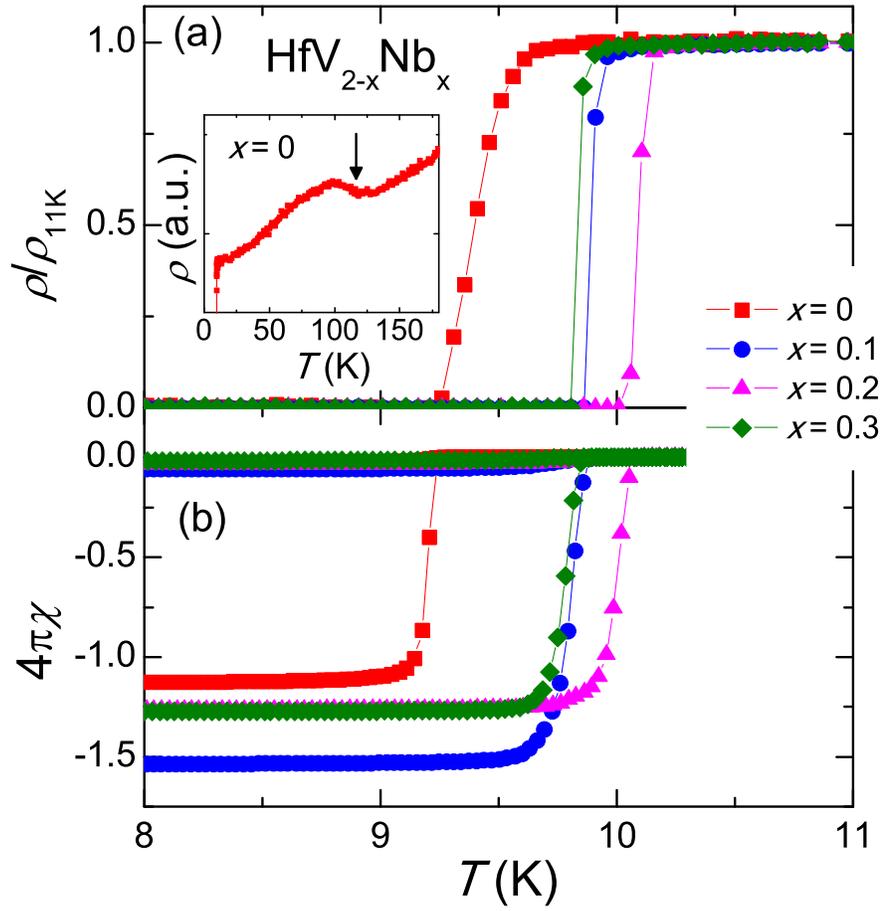}
\caption{\label{label} Temperature dependence of (a) resistivity and (b) magnetic susceptibility near the superconducting transition for the HfV$_{2-x}$Nb$_{x}$ samples.
The resistivity data are normalized for better comparison.
The inset in (a) shows a magnification of the normal state resistivity data, and the anomaly due to the structural transition is marked by the arrow.}
\end{figure}

\begin{figure}[h]
\centering
\includegraphics[width=12cm]{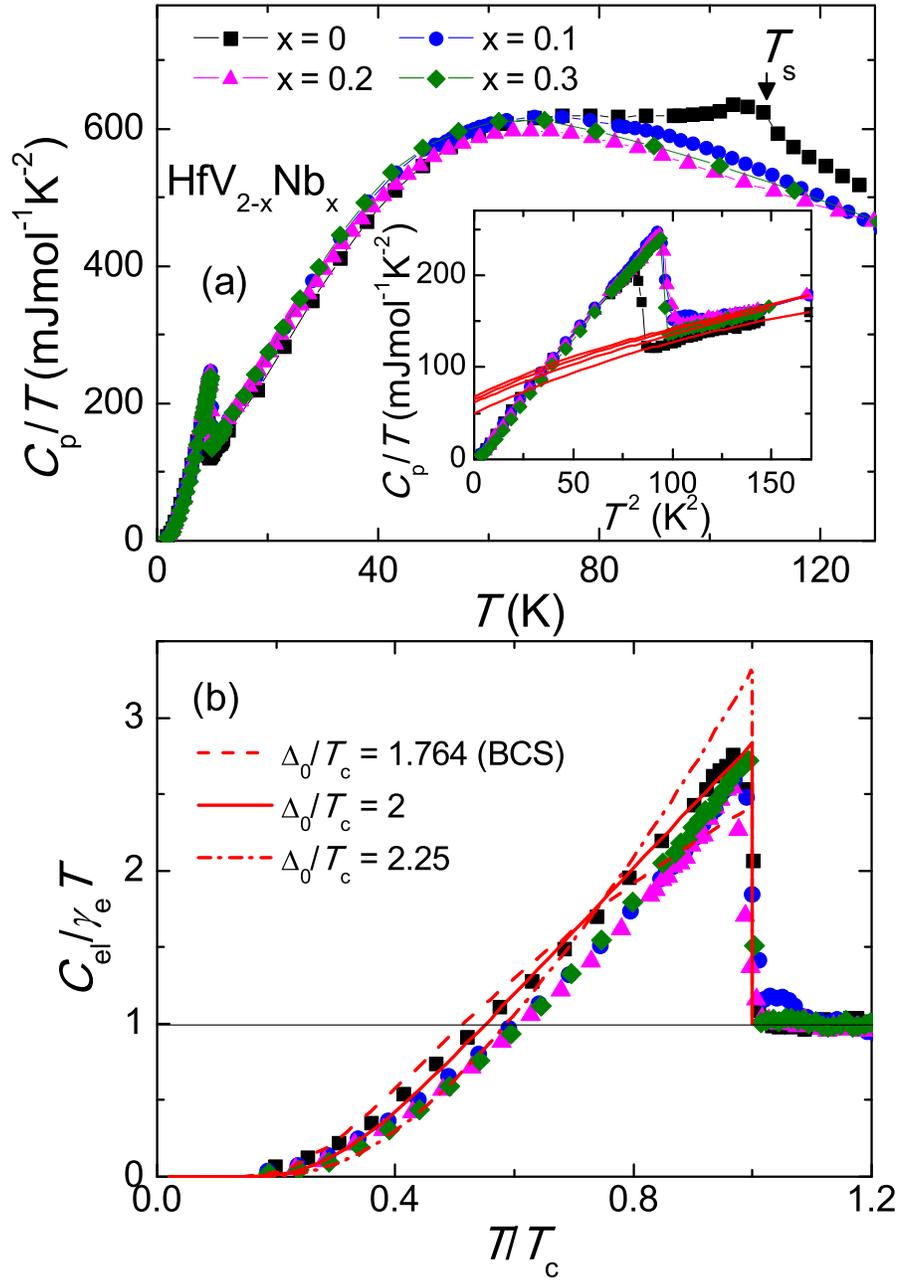}
\caption{\label{label} (a) Temperature dependence of specific heat ($C_{\rm p}$) for the HfV$_{2-x}$Nb$_{x}$ samples.
The anomaly around 110 K for $x$ = 0 is marked by the arrow.
The inset shows the low temperature data plotted as $C_{\rm p}$/$T$ versus $T^{2}$.
The solid lines are Debye fits to the normal-state data (see text for details).
(b) Normalized electronic specific heat plotted as a function of $T$/$T_{\rm c}$ for these samples.
The solid, dashed, and dashed dot lines denote the calculated $C_{\rm el}$/$T$ curves from the BCS theory and $\alpha$ model.}
\end{figure}

\begin{figure}[h]
\centering
\includegraphics[width=12cm]{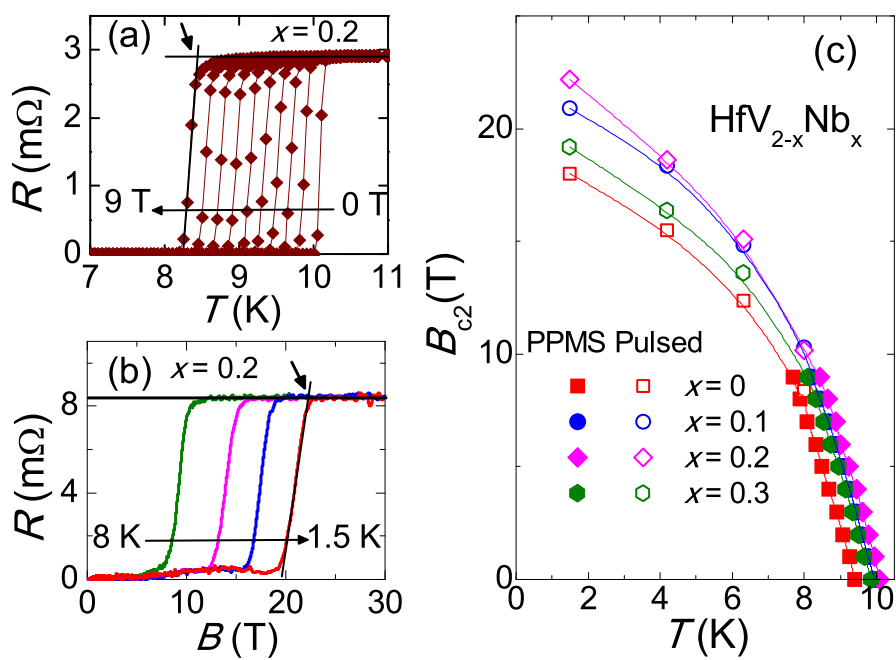}
\caption{\label{label} (a) Temperature dependence of resistivity for $x$ = 0.2 under various magnetic fields up to 9 T.
(b) Isothermal field dependence of resistivity up to 30 T for the same sample at different temperatures.
The solid lines and arrows are a guide to the eyes.
(c) Upper critical field $B_{\rm c2}$ versus temperature phase diagram for the HfV$_{2-x}$Nb$_{x}$ samples.}
\end{figure}

\begin{figure}[h]
\centering
\includegraphics[width=12cm]{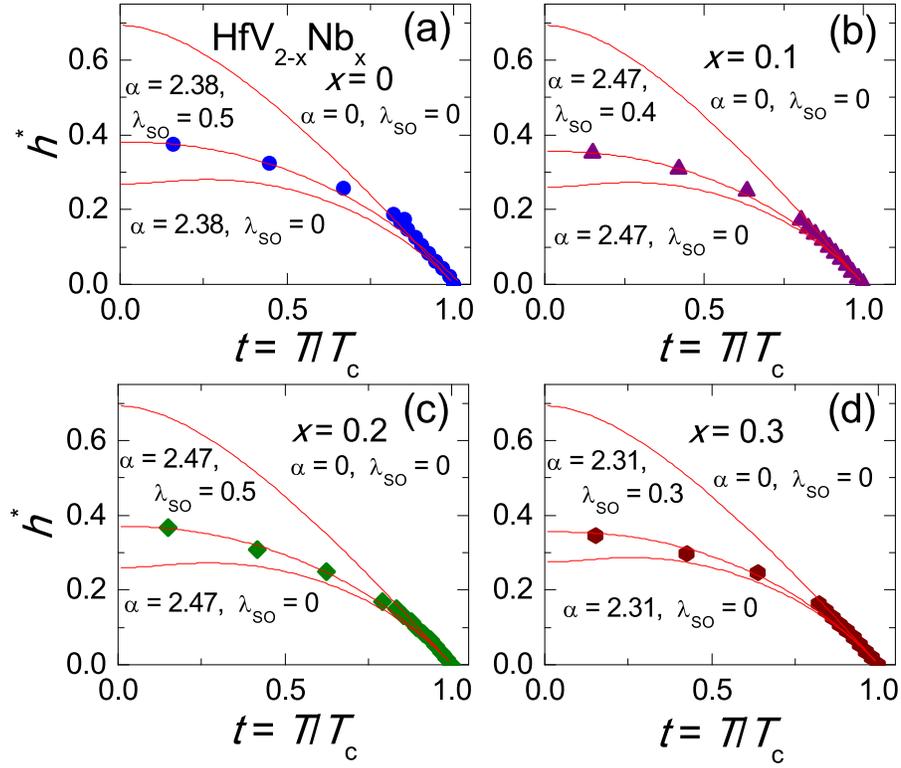}
\caption{\label{label} Normalized upper critical field $h^{\ast}$ plotted as a function of the reduced temperature $t$ = $T/T_{\rm c}$ for the HfV$_{2-x}$Nb$_{x}$ samples. (a) $x$ = 0; (b) $x$ = 0.1; (c) $x$ = 0.2; (d) $x$ = 0.3.
The solid lines in each panel denote the theoretical WHH curves in the absence of spin paramagnetic effect and spin-orbit scattering ($\alpha_{\rm M}$ = 0, $\lambda_{\rm SO}$ = 0), in the absence of spin-orbit scattering ($\alpha_{\rm M}$ $\neq$ 0, $\lambda_{\rm SO}$ = 0), and in the presence of both spin paramagnetic effect and spin-orbit scattering ($\alpha_{\rm M}$ $\neq$ 0, $\lambda_{\rm SO}$ $\neq$ 0). See text for details.}
\end{figure}

\begin{figure}[h]
\centering
\includegraphics[width=12cm]{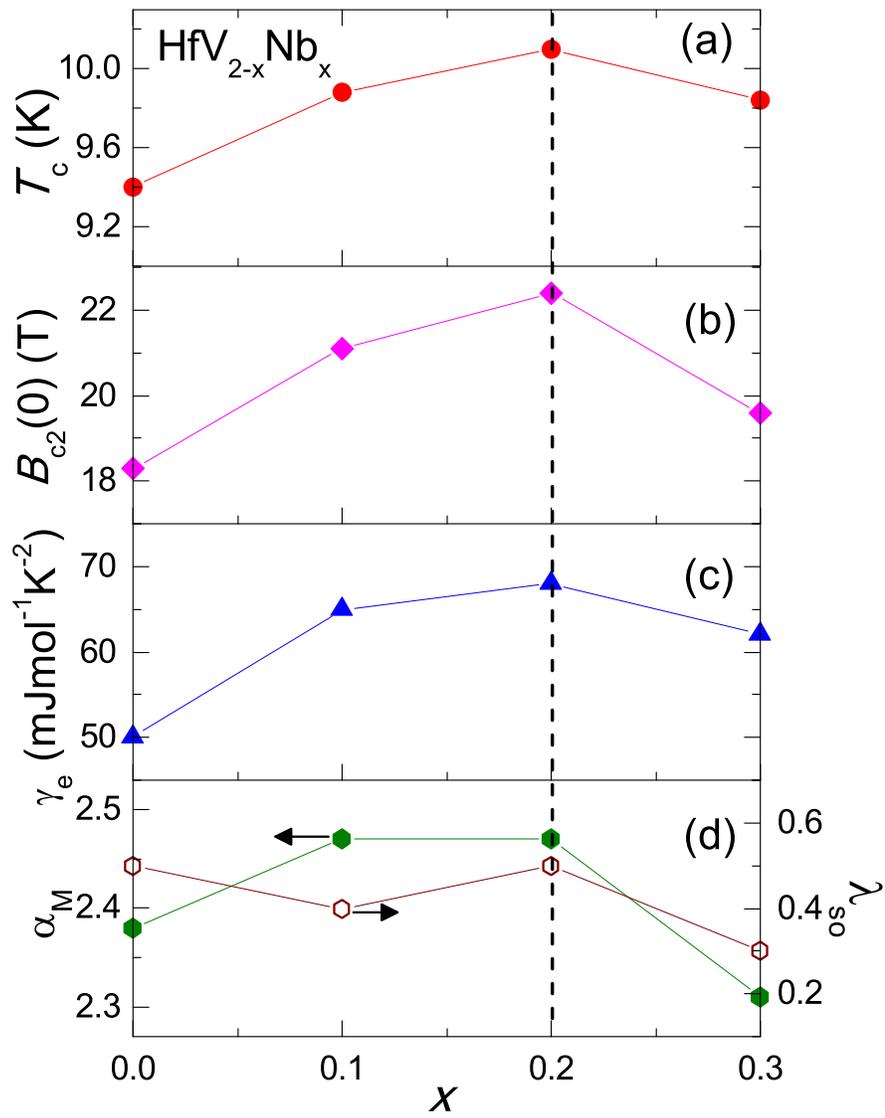}
\caption{\label{label} Nb content $x$ dependence of (a) $T_{\rm c}$, (b) $B_{c2}$(0), (c) $\gamma_{\rm e}$, and (d) $\alpha_{\rm M}$ and $\lambda_{\rm SO}$. The vertical line is a guide to the eyes.}
\end{figure}

\begin{figure}[h]
\centering
\includegraphics[width=12cm]{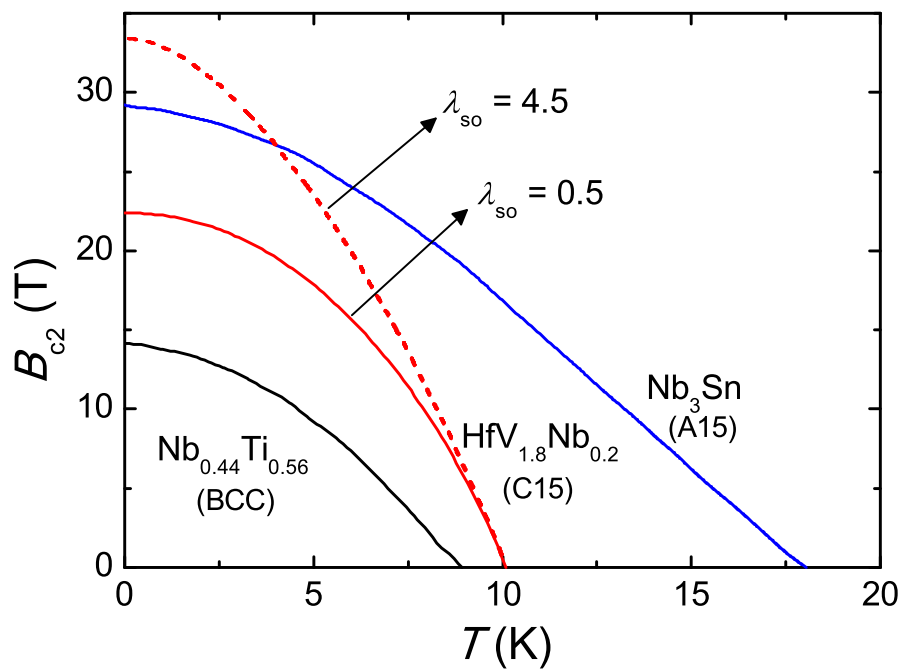}
\caption{\label{label} Comparison between the $B_{\rm c2}$-$T$ phase diagrams of HfV$_{1.8}$Nb$_{0.2}$, Nb$_{0.44}$Ti$_{0.56}$ \cite{Nb-Ti}, and Nb$_{3}$Sn \cite{Nb3Sn}.
The dashed line represents the theoretical curve for HfV$_{1.8}$Nb$_{0.2}$ assuming the same spin-orbit $\lambda_{\rm so}$ as Nb$_{0.44}$Ti$_{0.56}$.}
\end{figure}

\clearpage
\begin{table}
\caption{Structural and physical parameters of the HfV$_{2-x}$Nb$_{x}$ samples.}

\begin{indented}
\lineup
\item[]\begin{tabular}{@{}*{5}{l}}
\mr
 &$x$ = 0 &$x$ = 0.1 &$x$ = 0.2 &$x$ = 0.3 \\
\mr
$a$ ({\AA})							&  7.382 	 & 7.385	& 7.389 & 7.399 	\\
$\gamma_{\rm e}$ (mJmol$^{-1}$K$^{-2}$)							& 	  50 	 & 65 & 68& 62 		\\
$T_{\rm c}$	(K)			&      9.4    & 			9.88 & 10.1 & 9.84 	 \\
($d$$B_{\rm c2}/dT$)$_{T = T_{\rm c}}$  (T/K) 	&       $-$5.1 & $-$6.0 & $-$6.0 & $-$5.6   \\
$B_{\rm c2}^{\rm orb}$(0) (T)							& 	  33.2 	 & 41.1	& 41.9 & 38.2  	\\
$B_{\rm c2}^{\rm P}$(0) (T)							& 	  12.9 	 & 15.4 & 15.7 & 15.1 	\\
$B_{\rm c2}$(0) (T)				&      18.3    & 			21.1 & 22.4 & 19.6  	 \\
$\alpha_{\rm M}$ 	&       2.38 & 2.47 & 2.47 & 2.31  \\
$\lambda_{\rm so}$	&      0.5 &   0.4  & 0.5 & 0.3 \\
\br
\end{tabular}
\end{indented}
\end{table}

\clearpage
\begin{table}
\caption{Space group an superconducting parameters of HfV$_{1.8}$Nb$_{0.2}$, Nb$_{0.44}$Ti$_{0.56}$, and Nb$_{3}$Sn. The data for Nb$_{0.44}$Ti$_{0.56}$ and Nb$_{3}$Sn are taken from Ref. \cite{Nb-Ti} and \cite{Nb3Sn}, respectively.}

\begin{indented}
\lineup
\item[]\begin{tabular}{@{}*{4}{l}}
\br
 &Nb$_{0.44}$Ti$_{0.56}$ &\0Nb$_{3}$Sn &HfV$_{1.8}$Nb$_{0.2}$\\
\mr
\0\0Space group								 & \0\0 Im3m  &  \0 Pm3n & \0\0Fd3m		\\
\0\0$T_{\rm c}$ (K)					&\0\0 8.9	&  \0 17.6 & \0\0 10.1 \\
\0\0$B_{\rm c2}$(0)  (T)				&\0\0 14.2	& \0 28.9 &\0\0 22.4 \\
\0\0$B_{\rm c2}^{\rm orb}$(0)  (T)			 &\0\0 	  15.5	& \0 29.2 &\0\0 41.9	\\
\0\0$\alpha_{\rm M}$ 				 &\0\0 	  1.34	& \0 0 &\0\0	2.47 \\
\0\0$\lambda_{\rm so}$ 				 &\0\0 	  4.5	& \0 0 &\0\0	0.5 \\
\br
\end{tabular}
\end{indented}
\end{table}

\end{document}